\documentclass[prd,amsmath,amssymb,superscriptaddress,nofootinbib,twocolumn, letterpaper]{revtex4}

\usepackage{xcolor}
\usepackage{bm}
\usepackage[normalem]{ulem}
\usepackage{babel}
\usepackage{textcomp}
\usepackage{amstext}
\usepackage{graphicx}
\usepackage{esint}
\usepackage[unicode=true,pdfusetitle, bookmarks=true,bookmarksnumbered=false,bookmarksopen=false,
 breaklinks=false,pdfborder={0 0 1},backref=false,colorlinks=true]
 {hyperref}
 
\hypersetup{pdfborderstyle=,linkcolor=blue,citecolor=cyan}

\newcommand{\beq}{\begin{equation}}
\newcommand{\eeq}{\end{equation}}
\newcommand{\bqa}{\begin{eqnarray}}
\newcommand{\eqa}{\end{eqnarray}}

\newcommand{\bseq}{\begin{subequations}}
\newcommand{\eseq}{\end{subequations}}

\makeatletter

\usepackage{babel}
\usepackage{xcolor}
\usepackage{color}

\begin{document}

\title{Gluon Polarimetry with Energy-Energy Correlators}
 
\author{Yu-Kun Song}
\email{sps\_songyk@ujn.edu.cn }
\affiliation{School of Physics and Technology, University of Jinan, Jinan, Shandong 250022, China}

\author{Shu-Yi Wei}
\email{shuyi@sdu.edu.cn}
\affiliation{Institute of Frontier and Interdisciplinary Science, Key Laboratory of Particle Physics and Particle Irradiation (MOE), Shandong University, Qingdao, Shandong 266237, China}

\author{Lei Yang}
\email{lei.yang@mail.sdu.edu.cn}
\affiliation{Institute of Frontier and Interdisciplinary Science, Key Laboratory of Particle Physics and Particle Irradiation (MOE), Shandong University, Qingdao, Shandong 266237, China}

\author{Jian Zhou}
\email{jzhou@sdu.edu.cn}
\affiliation{Institute of Frontier and Interdisciplinary Science, Key Laboratory of Particle Physics and Particle Irradiation (MOE), Shandong University, Qingdao, Shandong 266237, China}
\affiliation{Southern Center for Nuclear-Science Theory (SCNT), Institute of Modern Physics, Chinese Academy of Sciences, HuiZhou, Guangdong
516000, China}

\begin{abstract}
We propose a novel method to probe gluon linear polarization via energy correlations in hard scattering processes. This approach exploits the characteristic $\cos 2\phi$ azimuthal modulation in single- and two-point energy correlations within jets initiated by polarized gluons. In contrast to conventional techniques that rely on $k_t$ resummation or intricate jet substructure observables, our method offers a theoretically robust and experimentally accessible avenue for gluon polarimetry. We perform an all-order analysis within the Ciafaloni-Catani-Fiorani-Marchesini (CCFM) formalism, incorporating coherent branching effects to achieve improved precision. Our predictions can be tested at current and future facilities, including the LHC, RHIC, HERA, and the EIC.
\end{abstract}

\maketitle

\date{\today}

{\it Introduction}--- Understanding the linear polarization of gluons inside unpolarized nucleons is essential for mapping their gluonic structure, a topic that has drawn significant interest from both small-$x$ and high-energy spin physics communities~\cite{Mulders:2000sh,Metz:2011wb,Schafer:2012yx,Boer:2009nc,Boer:2010zf,Qiu:2011ai,Pisano:2013cya,Boer:2016fqd,Boer:2017xpy,Kishore:2018ugo,Dominguez:2011br,Dumitru:2015gaa,Sun:2011iw,Boer:2011kf,Catani:2013tia,Boer:2014tka,Echevarria:2015uaa,Ma:2012hh,Gutierrez-Reyes:2019rug,denDunnen:2014kjo,Zhang:2014vmh,Ma:2015vpt,Altinoluk:2024vgg,Taels:2022tza}.  The recent finding~\cite{Bhattacharya:2024sno}, together with the observation that gluons exhibit 100\% linear polarization at small $x$ in the dilute limit~\cite{Metz:2011wb}, indicate that the helicity and orbital angular momentum of gluons are maximally quantum entangled. This phenomenon has broad phenomenological implications, as evidenced by predicted substantial $\cos 2\phi$ azimuthal asymmetries in two particle production processes~\cite{Boer:2009nc,Boer:2010zf,Qiu:2011ai,Pisano:2013cya,Boer:2016fqd,Schafer:2012yx,Boer:2017xpy} and the transverse momentum distribution of Higgs boson at the LHC~\cite{Sun:2011iw,Boer:2011kf,Catani:2013tia,Boer:2014tka,Echevarria:2015uaa,Ma:2012hh,denDunnen:2014kjo,Zhang:2014vmh,Ma:2015vpt,Gutierrez-Reyes:2019rug}. Despite these compelling signatures, directly measuring linear gluon polarization remains a major challenge. Traditional methods, such as those based on Transverse Momentum Dependent (TMD) factorization, suffer from two key limitations: the polarization signal is diluted by TMD evolution effects at low transverse momentum, and  identical  $\cos 2\phi$ asymmetries can also arise from final-state soft gluon emissions contaminating the extraction~\cite{Hatta:2021jcd,Hatta:2020bgy}. To address these challenges, emerging observables like Energy-Energy Correlations (EECs) offer a novel and potentially more precise method for measuring linear gluon polarization.

\begin{figure}[htb]
\centering
\includegraphics[width=0.95\linewidth]{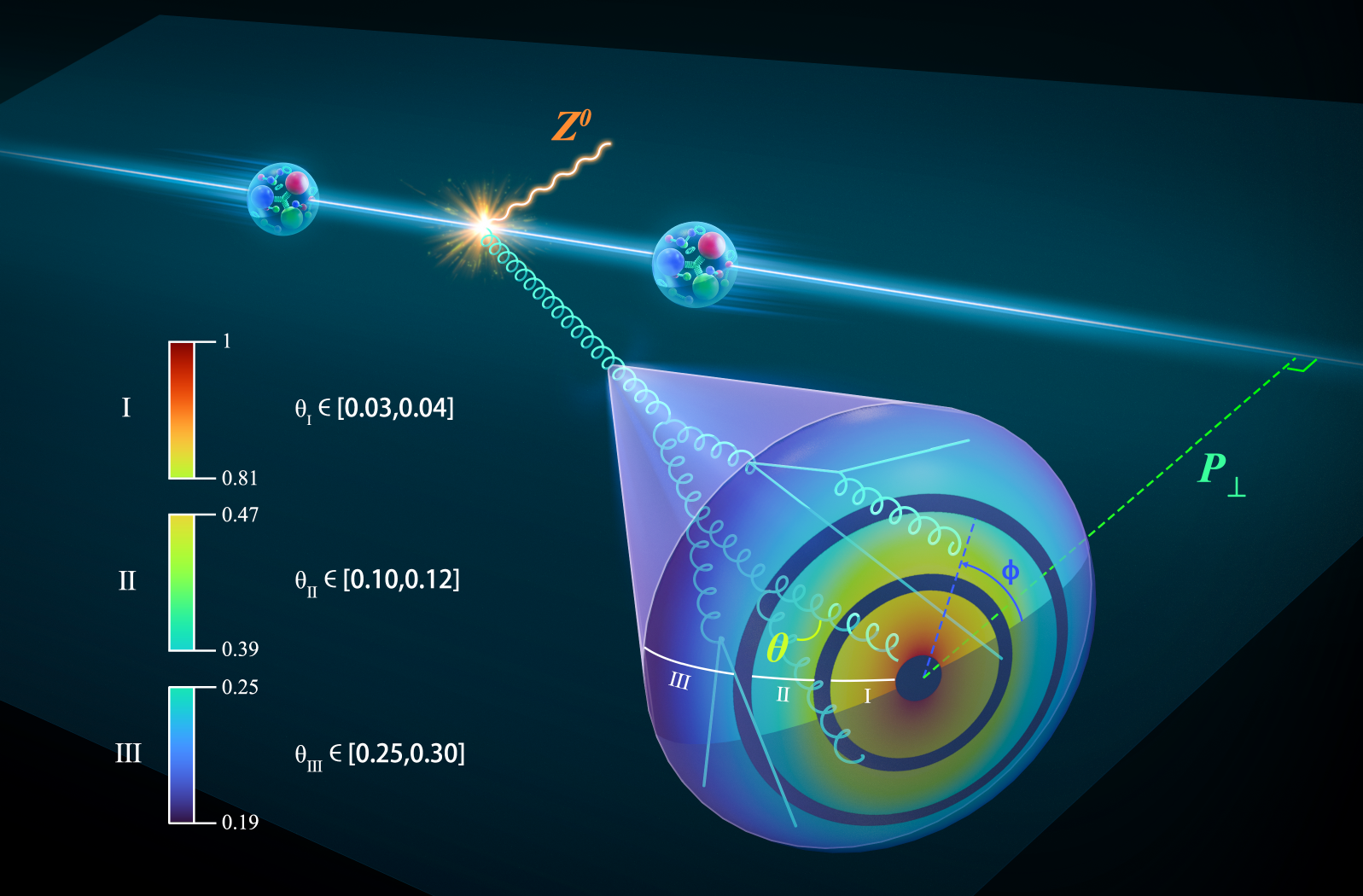}
\caption{Illustration of a proton-proton collision where a nearly collinear, linearly polarized gluon is emitted from an incoming parton in a hard scattering event. The gluon's transverse momentum $\vec{P}_\perp$ defines the reference axis. The magnified view shows the resulting EEC within the jet cone, displaying the $\cos 2\phi$ anisotropy, where $\phi$ is the angle between $\vec{P}_\perp$ and the detector pair orientation.}
\label{fig:illustration}
\end{figure}

EEC based methods also leverage the $\cos 2\phi$ azimuthal asymmetry to probe linear gluon polarization. Specifically, a $\cos 2\phi$ azimuthal modulation in the squeezed limit of the three-point energy correlator within a jet~\cite{Chen:2020adz,Chen:2021gdk} provides a direct probe of the linear polarization of intermediate virtual gluons.  Furthermore, the Nucleon EEC (NEEC) has emerged as a powerful  tool for studying linearly polarized gluon distributions~\cite{Li:2023gkh,Guo:2024jch}. 
The scale dependence of the near-side EEC observables is governed by the  DGLAP dynamics and is not sensitive to soft gluon radiations, yielding a theoretically cleaner analysis. Beyond jet substructure study, EEC, originally introduced in the 1970s~\cite{Basham:1978bw,Basham:1978zq} to test QCD in electron-positron colliders \cite{Schlatter:1981aw, CELLO:1982rca, Fernandez:1984db, TASSO:1987mcs, TOPAZ:1989yod, DELPHI:1990sof, OPAL:1991uui, Electron-PositronAlliance:2025fhk},  have evolved into a versatile precision tool with applications ranging from hot nuclear matter~\cite{Xing:2024yrb,Apolinario:2025vtx,Yang:2023dwc,Andres:2022ovj,Andres:2023xwr,Andres:2023ymw,Barata:2023bhh,Barata:2024nqo,Andres:2024ksi} to nucleon structure studies~\cite{Liu:2022wop,Guo:2024vpe,Guo:2024jch,Li:2023gkh,Liu:2023aqb,Bhattacharya:2025bqa,Mantysaari:2025mht,Kang:2023big}, and beyond~\cite{Ali:2012rn,Tulipant:2017ybb,Dixon:2019uzg,Karapetyan:2019fst,Luo:2019nig,Korchemsky:2019nzm,Moult:2019vou,Ebert:2020sfi,Gao:2020vyx,Li:2021txc,Li:2021zcf,Yu:2022kcj,Komiske:2022enw,Neill:2022lqx,Craft:2022kdo,ATLAS:2023lhg,Chen:2023wah,Devereaux:2023vjz,Schindler:2023cww,Chen:2023zlx,Kang:2023big,Cao:2023qat,ALICE:2024dfl,CMS:2024mlf,Jiang:2024qno,Aglietti:2024xwv,Singh:2024vwb,Chen:2024cgx,Shen:2024oif,Liu:2024lxy,Kang:2024dja,Gao:2024wcg,Fu:2024pic,Barata:2024wsu,Andres:2024xvk,Barata:2024ieg,Alipour-fard:2025dvp,Zhao:2025ogc,Kang:2025zto,Herrmann:2025fqy,Guo:2025zwb,Chang:2025kgq,Lee:2024esz,Lee:2025okn,Chen:2025rjc,Guo:2025qnz} (for a nice review, see Ref.~\cite{Moult:2025nhu} and references therein). Interestingly, Ref.~\cite{Chen:2024bpj} establishes  a robust connection between NEEC and fracture functions.

Inspired by recent developments~\cite{Chen:2020adz,Chen:2021gdk}, we introduce a novel approach to probe the linear polarization of collinear gluons emitted as initial-state radiation in $Z^0$ production process. As illustrated in Fig.~\ref{fig:illustration}, our method exploits the azimuthal dependence of two-point EEC measured within jets initiated by these gluons. The linear polarization vector is aligned with the gluon's transverse momentum vector. This polarization generates a characteristic $\cos 2\phi$ azimuthal asymmetry in EEC. Here, $\phi$ denotes the angle between the gluon's transverse momentum $\vec{P}_\perp$ and the vector connecting two detector pixels that contribute to the correlation. To achieve improved theoretical accuracy, we go beyond the conventional DGLAP formalism by incorporating coherent branching effects through the Ciafaloni-Catani-Fiorani-Marchesini (CCFM) framework~\cite{Ciafaloni:1987ur,Catani:1989yc,Catani:1989sg,Marchesini:1994wr}.  Notably, although partial cancellations occur between $g \rightarrow q\bar{q}$ and $g \rightarrow gg$ splittings, the resulting azimuthal asymmetry remains substantial. To enhance experimental accessibility, we further introduce a refined method to probe the linear polarization of gluon by employing the one-point energy correlation within the Winner-Takes-All (WTA) scheme, a well-established jet event shape observable~\cite{Bertolini:2013iqa,Neill:2016vbi,Kang:2017mda,Neill:2018wtk}. The polarization signature is effectively encoded in a similar $\cos 2\phi$ modulation, where $\phi$ is defined as the angle between $\vec{P}_\perp$ and the vector connecting the detector pixel to the WTA jet axis.  
  
Our methodology shares the strengths of other EEC-based methods, notably in avoiding the contamination from final state soft gluon radiations and eliminating the complexities introduced by Sudakov resummation in such analyses. Compared to prior EEC based techniques, such as those relying on three point correlations~\cite{Chen:2020adz,Chen:2021gdk} or the NEEC probe~\cite{Li:2023gkh}, our method enhances precision and experimental feasibility by directly targeting the asymmetry within jets produced via polarized gluon fragmentation. With this robust framework, we aim to set a new standard for gluon polarimetry.

\ 

{\it Formalism}---
Formally, the  differential two-point EEC inside a gluon jet can be defined as~\cite{Komiske:2022enw},
\begin{align}
 \frac{d\langle \text{EEC}\rangle}{d\theta\, d\phi} = \sum_{i,j} \delta(\theta_{ij} - \theta)\, \delta(\phi_{ij,g} - \phi)\, \frac{E_i E_j}{E_g^2},
\end{align}
where  the sum runs over all pairs of particles $(i, j)$ within the identified jet, and $E_g = \sum_i E_i$ is the total energy of the jet. Here, $\theta_{ij}$ is the angle between particles $i$ and $j$, and $\phi_{ij,g}$ denotes the azimuthal angle between the relative angle vector $\vec{\theta}_{ij}$ and the transverse momentum $\vec{P}_\perp$ of the jet.

To facilitate an all-order analysis, it is convenient to consider the $\theta$-integrated EEC, which can be expressed as a convolution of hard coefficients and the cumulant jet functions~\cite{Dixon:2019uzg}:
\begin{align}
\langle \text{EEC}\rangle =& \sum_{i=q,g}\int dx\, x^2H_i (x, \mu)\, J_i (\ln x^2 \kappa)
\end{align}
where $\kappa=\theta^2 E_g^2/\mu^2 $, and $x$ is the momentum fraction of the jet carried by the parton $i$. $J_i$ denotes the standard (unpolarized) cumulant jet function~\cite{Dixon:2019uzg}. $H_i$ is  the hard coefficient  of resolving a parton carrying momentum fraction $x$ within the gluon jet at the scale $\mu$, which will be specified later.

As discussed above, there is a direct correlation between the gluon branching plane and its polarization vector.  In the context of QCD evolution, this process can be naturally divided into two distinct stages:
First, as the parton shower develops, the linear polarization of gluons is progressively reduced due to scale evolution effects. Second, when a linearly polarized gluon with offshellness $E_g\theta$ undergoes a single splitting, the resulting energy distribution exhibits a characteristic azimuthal dependence.
 Taking these effects into account, we arrive at a factorized expression for the $\cos 2\phi$-weighted EEC:
\begin{align}
&\int \frac{d\phi}{\pi}\, \frac{d\langle \text{EEC}\rangle}{d\theta\, d\phi} \cos 2\phi = \int dx\, x^2 H_T(x, \mu)  \\
&\times \frac{\alpha_s}{2\pi} \frac{1}{\theta} \int \!dy y(1-y) \left[ P_{gg}^{2\phi}(y) + 2n_f P_{gq}^{2\phi}(y) \right] J_{g,T}(\ln x^2 \kappa)
\nonumber
\end{align}
Here, $J_{g,T}$ is the polarization-dependent jet function, which is the second moment of the jet function. Its operator definition is given by~\cite{Chen:2020adz}
\begin{align}
&\!\!\!\!\!\! J_{g,T}(\ln x^2 \kappa) = \frac{E_J}{2(N_c^2-1)} \sum_X \sum_{i,j \in X} \frac{E_i E_j}{E_g^2} \Theta(\theta_{ij} < \theta) \nonumber \\
&\times \langle 0 | \mathcal{B}_{\perp\mu}(0) | X \rangle \langle X | \mathcal{B}_{\perp\nu}(0) | 0 \rangle \left( g_\perp^{\mu\nu} + \frac{2 P_{\perp}^\mu P_{\perp}^\nu}{P_\perp^2} \right)
\end{align}
Here, $E_J$ is the energy of the jet.
The gauge-invariant gluon field strength operator is denoted by $\mathcal{B}_{\perp\mu}$. The projection tensor $(g_\perp^{\mu\nu} + 2 P_\perp^\mu P_\perp^\nu / P_\perp^2)$ isolates the linearly polarized component.
The relevant azimuthal-dependent splitting functions are:
\begin{align}
P_{gg}^{2\phi}(y) = 2C_A\, y(1-y), \quad 
P_{qg}^{2\phi}(y) = -y(1-y)
\end{align}
$H_T$ is the corresponding hard coefficient for producing a linearly polarized gluon.

To proceed further, we note that the observable studied here is closely related to the NEEC~\cite{Liu:2022wop,Li:2023gkh,Guo:2024jch}, since we focus on gluon radiation in the forward region along the beam direction, where $P_\perp^2 \ll Q^2$. We therefore adopt a similar factorization approach, in which the initial-state collinear gluon radiation is separated from the subsequent hard scattering process. As an example, consider $Z^0$ boson production, where an incoming quark emits a gluon before the hard interaction:
\begin{align}
\frac{d\sigma_U}{d\xi dy_Z d^2P_\perp} &= \sigma_0\, \frac{\alpha_s}{2\pi}\, C_F\, \frac{1}{P_\perp^2}\, \frac{1 + (1-\xi)^2}{\xi}, \\
\frac{d\sigma_T}{d\xi dy_Z d^2P_\perp} &= \sigma_0\, \frac{\alpha_s}{2\pi}\, C_F\, \frac{1}{P_\perp^2}\, \frac{2(1-\xi)}{\xi},
\end{align}
where $\sigma_0$ is the Born-level cross section for $Z^0$ production, and $\xi$ denotes the longitudinal momentum fraction carried by the gluon jet. The ratio $\sigma_T/\sigma_U$ quantifies the degree of linear polarization of the emitted gluon. Therefore, a natural choice for hard coefficients reads $H_i(x, \mu = \theta_0 E_g) = \delta_{ig} \delta(x-1)$ and $H_T(x,\mu = \theta_0 E_g) = \sigma_T/\sigma_U \delta (x-1)$, with $\theta_0 \sim {\cal O}(1)$.

In the DGLAP formalism, the evolution equations of the jet functions are given by
\begin{align}
& \frac{\partial  J_{i} (\ln \kappa)}{\partial \ln \mu^2} = \frac{\alpha_s}{2\pi} \sum_{j} \int_0^1 dy y^2  P_{ij} (y) J_{j} (\ln (y^2 \kappa) ),
\\
& \frac{\partial  J_{g,T} (\ln \kappa)}{\partial \ln \mu^2} = \frac{\alpha_s}{2\pi}  \int_0^1 dy y^2  P_{gg}^{T} (y) J_{g,T} (\ln (y^2 \kappa) ).
\end{align}
  Here $i,j$ denote the parton flavors, and $P_{ij}$ is the unpolarized splitting function.
 $P_{gg}^{T} (y)$ is the linearly polarized gluon splitting function, which at the LO reads~\cite{Sather:1990bq,Ma:2013yba}
\begin{equation}
P^T_{gg} (y)= 2C_A \frac{y}{(1 - y)_+}+\beta_0 \delta(1-y),
\end{equation}
with $\beta_0 = (33-2n_f)/6$.

Having discussed two-point EECs, we now consider the single-point energy correlation, or jet event shape.  This observable quantifies the energy distribution within a jet by measuring the correlation between the energy of a radiated parton and the jet core itself.  A key advance in this context is the Winner-Take-All (WTA) jet axis reconstruction scheme~\cite{Bertolini:2013iqa,Neill:2016vbi,Kang:2017mda,Neill:2018wtk}. Unlike conventional algorithms that define the jet axis as the sum of all constituent momenta, the WTA algorithm assigns the jet axis to the direction of the hardest particle at each recombination step. This procedure effectively eliminates recoil effects from soft gluon radiation, resulting in a recoil-free jet axis that enables more precise measurements of jet substructure and polarization-sensitive observables.

Within the WTA scheme, the unpolarized and linearly polarized single-point jet functions are denoted by $D_{g/q}$ and $D_{g,T}$, respectively. In the DGLAP formalism, their scale evolution is governed by
\begin{align} 
& \frac{\partial D_{i} (\ln \kappa)}{\partial \ln  \mu^2 } = \frac{\alpha_s}{2\pi} \sum_j \int_{\frac{1}{2}}^1 dy y P_{ij}(y)  D_{j} ( \ln y^2 \kappa ),
\\
& \frac{\partial D_{g,T} (\ln \kappa)}{\partial \ln  \mu^2 } = \frac{\alpha_s}{2\pi}  \int_{\frac{1}{2}}^1 dy y P_{gg}^{T}(y)  D_{g,T} ( \ln y^2 \kappa ).
\end{align}
Here, the lower integration limit $y=1/2$ arises from the kinematic constraint in the WTA scheme. Additionally, note that the evolution equation involves a single power of $y$, in contrast to the two-point EEC case where $y^2$ appears. Note that the one point energy correlation under investigation differs from that studied in Ref.~\cite{Mi:2025abd}.

While the DGLAP formalism serves as the standard approach for describing the scale evolution of jet functions, its conventional implementation based on virtuality ordering does not intrinsically account for the destructive interference of soft gluons at large angles. The CCFM formalism~\cite{Ciafaloni:1987ur,Catani:1989yc,Catani:1989sg,Marchesini:1994wr} addresses this by implementing angular ordering: each successive emission in the final-state parton shower must occur at a smaller opening angle than the previous one. This approach not only regulates infrared divergences more effectively, ensuring that the EEC remains finite as the angular separation $\theta \to 0$, but also yields a more realistic description of energy flow and correlations within jets. 

In the original CCFM approach, it is necessary to introduce a parton distribution that depends on the transverse momentum in order to resum large logarithms at small $x$. By analogy, we also define a transverse momentum dependent jet function, $J_g(\ln \kappa, k_t)$, which satisfies the normalization condition $\int d^2k_t\, J_g(\ln \kappa, k_t) = J_g(\ln \kappa)$. This construction allows us to directly apply the CCFM formalism to the evolution of the jet functions.

Following Refs.~\cite{Marchesini:1994wr,Jung:2001hx,CASCADE:2010clj}, and focusing for simplicity on the gluon-to-gluon channel, the CCFM evolution equation for the unpolarized gluon jet function can be written in differential form as:
\begin{align}
&\frac{\partial}{\partial \ln \mu^2} \frac{J_g(\ln \kappa, k_t)}{\Delta_s(\mu^2)} = \frac{\alpha_s}{2\pi} \int_{\Lambda/\mu}^{1-\Lambda/\mu} dy\, y^2 
\\
&\qquad \times \int_0^{2\pi} \frac{d\phi'}{2\pi} \frac{\tilde{P}_{gg}(y, \mu^2, k_t)}{\Delta_s(\mu^2)} J_g(\ln \kappa, |\vec{k}_t + (1-y)\vec{\mu}|),\nonumber
\end{align}
where $\mu^2 = \vec{\mu}^2$ sets the maximum allowed angle for gluon emission, $\theta_{\text{max}} = \mu/E_g$, and $\Lambda$ is an infrared cutoff. The angle $\phi'$ denotes the azimuthal separation between $\vec{k}_t$ and $\vec{\mu}$. The Sudakov form factor $\Delta_s$ resums the effects of virtual and unresolved real emissions. Importantly, due to the implementation of angular ordering in CCFM, the $y$-dependence in the logarithm is eliminated.

In the CCFM framework, the splitting function $\tilde{P}_{gg}$ is unregularized, retaining only the singular terms  $1/y$ and $1/(1-y)$. In the original context, a non-Sudakov form factor $\Delta_{ns}$ is introduced to resum logarithms of the form $\ln(1/y)$. However, for energy-energy correlation (EEC) observables, the energy-weighting factors $y$ or $y^2$ suppress the $1/y$ divergence, making small-$x$ resummation via the non-Sudakov factor unnecessary. Consequently, the unregularized splitting function for the unpolarized case simplifies to $\tilde{P}_{gg}(y) = 2 C_A \left[ \frac{y}{1-y} + \frac{1-y}{y} + y(1-y) \right]$, which is independent of the transverse momentum $k_t$.

With this simplification, we can integrate over $k_t$ to obtain an evolution equation for the inclusive jet function:
\begin{align}
\frac{\partial}{\partial \ln \mu^2} \frac{J_g(\ln \kappa)}{\Delta_s(\mu^2)} = \frac{\alpha_s}{2\pi} \int_{\Lambda/\mu}^{1-\frac{\Lambda}{\mu}} \!\!\! dy y^2\tilde{P}_{gg}(y)
\frac{J_g(\ln \kappa)}{\Delta_s(\mu^2)}.
\end{align}
The corresponding Sudakov form factor is given by
\begin{align}
&  \Delta_s(\mu^2)=\\
& \exp\Bigg\{ \!- \!\int_{4\Lambda^2}^{\mu^2} \!\! \frac{d\mu'^2}{\mu'^2} \frac{\alpha_s}{4\pi}\int_{\Lambda/\mu'}^{1-\frac{\Lambda}{\mu'}} \!\! dy  [\tilde{P}_{gg}(y)+ 2n_f \tilde{P}_{qg}(y)] \Bigg\}.  \nonumber 
\end{align}
where $\tilde{P}_{qg}(y)$ is the unregularized splitting function for the gluon-to-quark channel.
This form closely parallels the Sudakov factor used in parton shower algorithms.

Extending the above discussion to the case of linearly polarized gluons, we find that the evolution of the corresponding jet function $J_{g,T}$ within the CCFM framework takes a similar form to the unpolarized case, 
\begin{align}
\frac{\partial }{\partial \ln \mu^2} \frac{J_{g,T}(\ln \kappa )}{\Delta_s(\mu^2)} = \frac{\alpha_s}{2\pi} \int_{\Lambda/\mu}^{1-\frac{\Lambda}{\mu}} \!\!\! dy\, y^2 
 \frac{\tilde{P}_{gg}^T(y)}{\Delta_s(\mu^2)}  J_{g,T}(\ln \kappa),
\end{align}
where the unregularized splitting function for linearly polarized gluons is given by $\tilde{P}_{gg}^T(y) = 2C_A \frac{y}{1-y}$. 

The CCFM formalism can be directly extended to single-point energy correlations in the WTA scheme. In the next section, we examine the phenomenological consequences of these improvements, particularly for  azimuthal asymmetries from linearly polarized gluons.

\ 

{\it Phenomenological Analysis}--- We now turn to the phenomenological analysis of EECs and their sensitivity to gluon polarization.

The boundary conditions for the jet functions are set at a reference angle $\theta_0$ (e.g., $\theta_0 = 0.3$), corresponding to a factorization scale $\mu_0 = P_\perp = 30$~GeV for gluon energy $E_g = 90$~GeV, as motivated by the associated $Z^0$ production process. Specifically, we impose
\begin{align}
J_g(0) = 1, \  J_q(0) = 0, \ J_{g,T}(0) = 1,
\end{align}
The quark jet function is initialized to zero, reflecting the fact that quark contributions arise only from subsequent branchings.

Collecting all pieces together, the azimuthal asymmetry for  the EEC, which encodes the effect of gluon linear polarization, is given by
$ \langle \cos 2\phi \rangle = A(\theta) H_T / H $
where   the analyzing power $A(\theta)$ is defined as
\begin{align}
A(\theta) \equiv\frac{
 \int \! dy y(1-y) \left [ P_{gg}^{2\phi}+2n_f P_{qg}^{2\phi} \right ] \! J_{g,T}
}{
  \int \! dy y(1-y) \! \left \{ \left [ \tilde P_{gg}+2 n_f \tilde P_{qg }\right ]  \! J_{g}  + \! \left [ \tilde P_{qq}+ \tilde P_{gq }\right ] \!  J_{q} \right \}
},
\end{align}
with the limits of the $y$ integration being set to $\int_{\Lambda/\mu_0}^{1-\Lambda/\mu_0}$ in order to account for the coherent branching effects. For the single-point energy correlation in the WTA scheme, the expression  of the asymmetry remains analogous. The analyzing power $A(\theta)$ serves as a key measure of the observable's sensitivity to gluon polarization, and is central to our phenomenological study.

\begin{figure}[h!]
\centering
\includegraphics[width=0.9\linewidth]{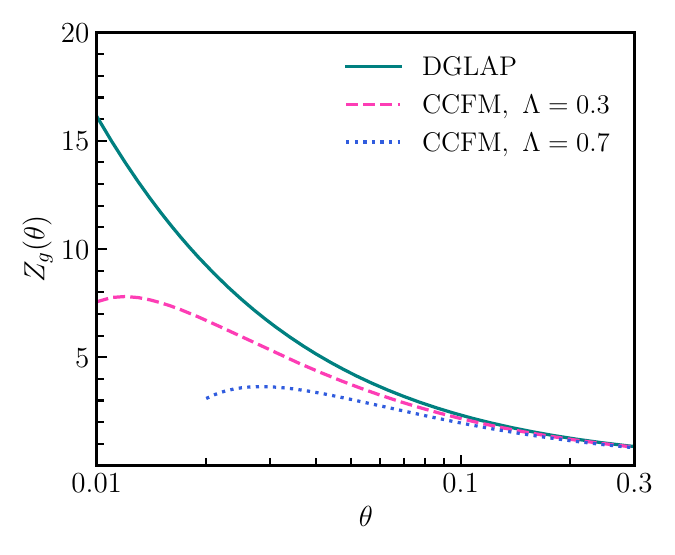}
\caption{Normalized unpolarized angular distributions for the EEC in the region $\theta\in [0.01, 0.3]$, calculated with the boundary conditions $J_{g}(0) = 1$ and $J_{q}(0) = 0$ at $\mu= 30$~GeV. Results are shown for both standard DGLAP evolution and the CCFM formalism with infrared cutoffs $\Lambda = 0.3$~GeV and $\Lambda =0.7$~GeV.}
\label{fig:unp}
\end{figure}

We now present numerical results for  the unpolarized distribution and the $\cos 2\phi$ azimuthal modulations. Figure~\ref{fig:unp} shows the self-normalized unpolarized angular distributions for the EEC: $Z_g(\theta) = \partial J_g(\ln\kappa)/ \partial \theta$  in the region $\theta \in [0.01, 0.3]$, using the boundary conditions specified above. We compare results obtained from standard DGLAP evolution with those from the CCFM formalism, using infrared cutoffs $\Lambda = 0.3$~GeV and $\Lambda = 0.7$~GeV. Notably, our calculations qualitatively reproduce the characteristic plateau behavior of EEC. This plateau signifies the onset of confinement and the transition to the non-perturbative regime. The CCFM framework, by incorporating angular ordering and an effective infrared cutoff, successfully captures the dynamics of this transition, which remains challenging for traditional collinear factorization. A comparison between theoretical calculations in the DGLAP and CCFM formalisms and data from $e^+e^-$-annihilation experiments is also presented in the Supplemental Material \cite{SuppMate}.

\begin{figure}[h!]
\centering
\includegraphics[width=0.9\linewidth]{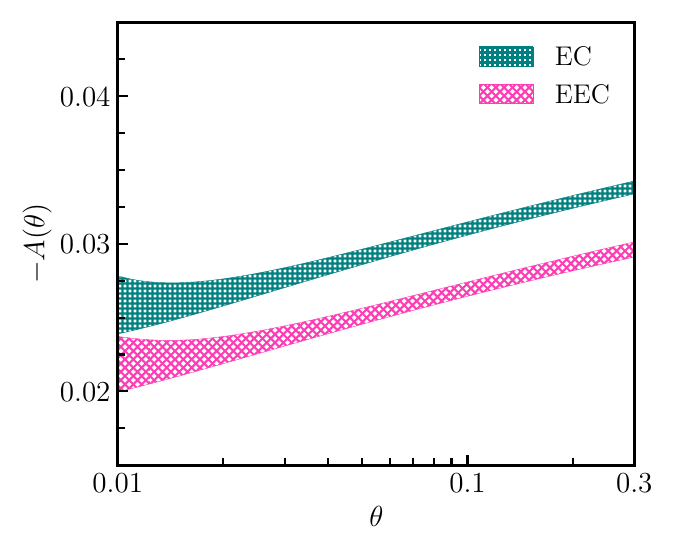}
\caption{Gluon polarization analyzing power $A(\theta)$, representing the maximal azimuthal asymmetry $\langle \cos{2\phi} \rangle$, for single-point  and two-point energy correlations with $n_f = 5$, plotted as a function of $\theta$ in the range $0.01$ to $0.3$ at $\mu = P_\perp= 30$~GeV. The error bands show the effect of varying the infrared cutoff $\Lambda$ from $0.3$~GeV to $0.7$~GeV.}
\label{asymmetry}
\end{figure}

Figure~\ref{asymmetry} illustrates the analyzing power $A(\theta)$, which quantifies the maximal $\cos 2\phi$ azimuthal asymmetry, for both single-point and two-point energy-energy correlations (EECs) as a function of the angle $\theta$ in the range $0.01$ to $0.3$. The shaded error bands indicate the impact of varying the infrared cutoff $\Lambda$ between $0.3$~GeV and $0.7$~GeV. Despite the sensitivity of the unpolarized distribution to $\Lambda$ in the small-angle region, the analyzing power $A(\theta)$ remains remarkably robust. This stability arises from the cancellation of common non-perturbative effects in the ratio, confirming that the polarization signal is a resilient observable. The results demonstrate that the azimuthal asymmetry is sizable and should be accessible to measurement at the LHC.

Recent ATLAS~\cite{ATLAS:2023lhg} and CMS~\cite{CMS:2024mlf} measurements demonstrate the high precision achievable in mapping the radial dependence of jet EECs. Our proposal extends these studies by retaining the azimuthal $(\phi)$ correlation relative to the jet $\vec{P}_\perp$. Since $\vec{P}_\perp$ is a standard observable, extracting the predicted $\cos 2\phi$ modulation requires only a modified binning strategy rather than new detector capabilities. This makes existing LHC data sets ideal for seeking direct evidence of linearly polarized gluons.

Moreover, the potential cancellation of the $\cos 2\phi$ modulation between $g \to q\bar{q}$ and $g \to gg$ splittings can be effectively bypassed through flavor-tagging techniques. For instance, the analyzing power is significantly enhanced by isolating the $g \to b\bar{b}$ or $g\to c\bar c$ channel. Recently, the CMS collaboration \cite{CMS:2025awh} also developed a new technique to select $g\to q\bar q$ branching events, and reported the first direct observation of spin correlations \cite{Collins:1987cp} induced by gluon polarization within jets. To illustrate the effectiveness of flavoring tagging, we perform a fixed-order calculation of the $\cos 2\phi$ azimuthal asymmetry for a $c\bar{c}$ pair produced within a gluon-initiated jet in $Z+\text{jet}$ production at the Tevatron ($\sqrt{s}=1.96\,\mathrm{TeV}$) utilizing scattering matrix elements presented in Refs.~\cite{Ellis:1985er, Campbell:2000bg, Harris:2001sx, Owens:2001rr}. These cuts reveal a large modulation reaching $\sim \mathbf{40\%}$, demonstrating the significant sensitivity enhancement provided by heavy-flavor tagging. More details are provided in the Supplemental Material \cite{SuppMate}.

\

{\it Summary}--- We introduce a novel method to probe the linear polarization of gluons by analyzing the energy distribution within jets produced by initial-state radiation in Drell-Yan-like processes.   Gluon polarization appears as a characteristic $\cos 2\phi$ modulation in the azimuthal dependence of the EEC and jet event shapes within the WTA scheme.

\begin{figure}[htb]
\centering
\includegraphics[width=0.95\linewidth]{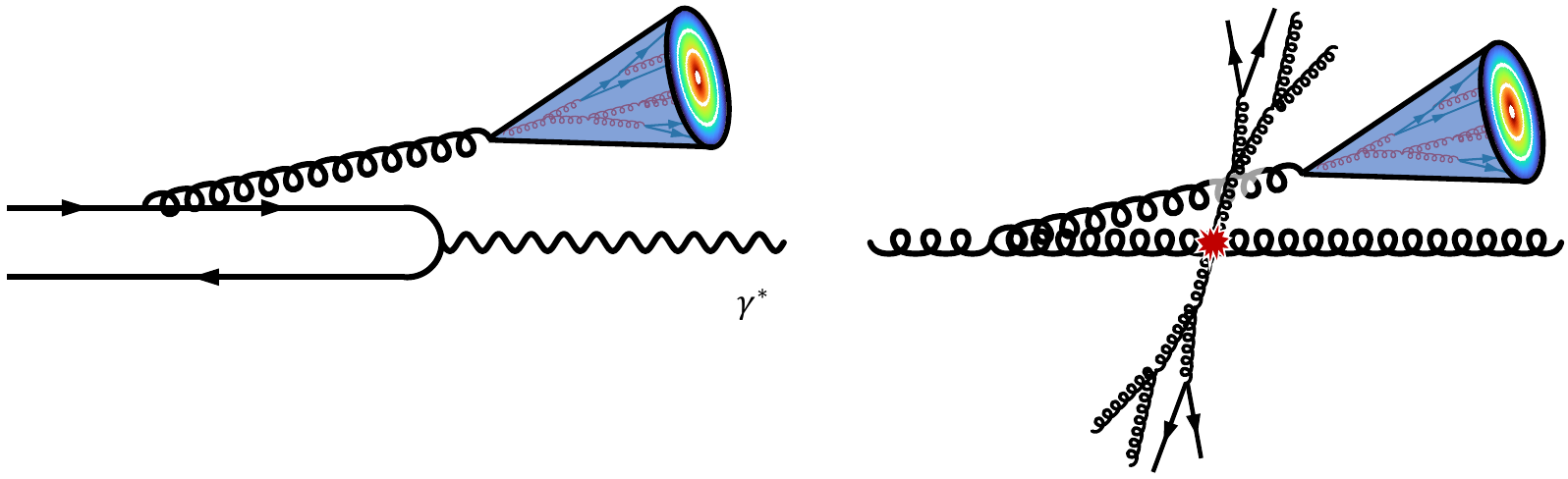}
\caption{Schematic illustration of linearly polarized gluons from initial-state radiation in DIS (left) and di-jet production (right), showing characteristic energy patterns in jet cones that enable probing gluon polarization.}
\label{fig:gamma_q-gg}
\end{figure}

Our approach is innovative both phenomenologically, as it provides direct sensitivity to gluon polarization without relying on complex declustering or tagging algorithms, and theoretically, through the use of the CCFM formalism to incorporate coherent branching effects. Notably, the CCFM formalism naturally captures the plateau behavior observed in the EEC, providing a unified description of both perturbative and nonperturbative effects.  While our study centers on Drell-Yan processes, the technique extends seamlessly to other hard scattering processes, such as DIS and mid-rapidity di-jet production accompanied by initial-state gluon radiation as shown in Fig.~\ref{fig:gamma_q-gg}. The predicted asymmetries can be validated with existing data from HERA, RHIC, and the LHC, with further tests anticipated at the future EIC. In summary, this approach establishes a new framework for gluon polarimetry with EECs in the perturbative large-angle regime, while shedding new light on the transition from non-perturbative to perturbative collinear dynamics in the polarization-dependent case.

{\it Acknowledgements:}
We thank Xiao-hui Liu, Ting Lin, Jin-long Zhang, and Wen-Bin Zhao for helpful discussions. This work is supported by the National Science Foundations of China under Grant No.~12575091, No.~12405156, No.~12321005, No.~12175118, and No.~11505080, the Shandong Province Natural Science Foundation under grant No.~2023HWYQ-011, No.~ZFJH202303 and No.~ZR2018JL006, and the Taishan fellowship of Shandong Province for junior scientists. 

{\it Data Availability:} 
The data that support the findings of this article are openly available \cite{DATAFILE}.


\clearpage

\onecolumngrid 

\setcounter{equation}{0}
\setcounter{figure}{0}
\setcounter{table}{0}

\renewcommand{\thesection}{S-\Roman{section}}  
\renewcommand{\theequation}{S\arabic{equation}}
\renewcommand{\thefigure}{S\arabic{figure}}
\renewcommand{\thetable}{S\arabic{table}}

\begin{center}
    \textbf{\large Supplemental Material} \\[0.5cm]
    Yu-Kun Song, Shu-Yi Wei, Lei Yang, Jian Zhou\\[0.4cm]
\end{center}


\begin{list}{}{
    \setlength{\leftmargin}{1.9cm}  
    \setlength{\rightmargin}{1.9cm} 
}
\item[] 
\small
In this supplemental material, we provide additional phenomenological results supporting the findings of our paper. In Sec.I, we derive the factorized formalism for the gluon linear polarization in the $q+\bar q\to g+Z^0$ process and numerically investigate the corrections in terms of $P_\perp^2/M_Z^2$. In Sec. II, we present a comparative analysis of the EEC in $e^+e^-$ annihilation using both DGLAP and CCFM formalisms. We demonstrate that the CCFM approach, by incorporating angular ordering and an effective infrared cutoff, successfully reproduces the plateau behavior observed in experimental data at small angles. In Sec. III, we present a fixed-order calculation of the EEC for charm-anticharm pairs in $p\bar{p}$ collisions. We show that while inclusive jet production exhibits negligible azimuthal asymmetry, $Z^0$-tagged jet events display a significant $\cos 2\phi$ modulation, reaching approximately 40\%, which confirms the enhanced sensitivity of heavy-flavor tagged observables to gluon linear polarization.
\end{list}





\section{Gluon linear polarization in the $q + \bar q \to g + Z^0$ process}

The factorized framework employed in this work assumes that the gluon $P_\perp$ is small compared to the hard scale $Q$ (e.g., the dilepton invariant mass in Drell-Yan or the $Z^0$ mass), such that $P_\perp^2/Q^2 \ll 1$. In this limit, the gluon can be treated as collinearly radiated from the initial state. Quantitatively, our approximation is generally valid for $P_\perp \simeq 0.3 Q$. For $Z^0$ production ($Q = M_Z \approx 91$ GeV), this corresponds to $P_\perp \simeq 30$ GeV. 

To quantify the sensitivity of our predictions to finite-$P_\perp$ corrections, we compare the exact partonic cross sections with the collinear asymptotic expressions. The exact unpolarized ($\sigma^U$) and polarized ($\sigma^T$) cross sections for the $q + \bar q \to g + Z^0$ channel are given by, 
\begin{align}
&
\frac{1}{\pi} \frac{d\hat\sigma^U_{q\bar q\to g+Z^0}}{d\hat t} = \frac{8 \alpha_s \alpha_{\rm em} c_1^q}{9 \sin^2 2\theta_W} \frac{1}{\hat s^2} \left( \frac{2M_Z^2 \hat s}{\hat u \hat t} + \frac{\hat u}{\hat t} + \frac{\hat t}{\hat u} \right),
\\
&
\frac{1}{\pi} \frac{d\hat\sigma^T_{q\bar q\to g+Z^0}}{d\hat t} = \frac{8 \alpha_s \alpha_{\rm em} c_1^q}{9 \sin^2 2\theta_W} \frac{1}{\hat s^2} \frac{2 M_Z^2 \hat s}{\hat u \hat t},
\end{align}
where $c_1^q = (c_V^q)^2+(c_A^q)^2$ represents the $Z^0 q\bar q$ coupling, $\theta_W$ is the Weinberg angle, and $\hat s$, $\hat t$, and $\hat u$ are the standard Mandelstam variables. Consequently, the gluon linear polarization reads
\begin{align}
P_g^T (\hat s, \hat t, \hat u) = \frac{d\hat \sigma^T /d\hat t}{d\hat \sigma^U/d\hat t} = \frac{2M_Z^2 \hat s}{2M_Z^2 \hat s + \hat u^2 + \hat t^2}.
\label{eq:gluon_pol}
\end{align}

Without loss of generality, we assume the gluon is produced in the forward rapidity region in the $q\bar q$ center-of-mass frame, and define $\xi$ as the momentum fraction of the quark carried by the gluon. The Mandelstam variables can be expressed in terms of $\xi$ and $P_\perp$ as
\begin{align}
& 
\hat s = \frac{P_\perp^2 + \xi M_Z^2}{\xi (1-\xi)},
&&
\hat t = - \frac{P_\perp^2}{\xi},
&&
\hat u = - \frac{P_\perp^2 + \xi M_Z^2}{1-\xi}.
\end{align}
The gluon linear polarization can thus be expressed as a function of two variables: $\xi$ and $P_\perp$.

In the limit $P_\perp^2 \ll M_Z^2$, the partonic cross sections reduce to the asymptotic forms
\begin{align}
&
\frac{1}{\pi} \frac{d\hat\sigma^U_{q\bar q\to g+Z^0}}{d\hat t} \simeq 
\frac{8 \alpha_s \alpha_{\rm em} c_1^q}{9 \sin^2 2\theta_W} \frac{1}{\hat s} \frac{1}{P_\perp^2} [1+(1-\xi)^2] = \sigma_0^q \frac{1}{P_\perp^2}\frac{\alpha_s}{2\pi} C_F [1+(1-\xi)^2],
\\
&
\frac{1}{\pi} \frac{d\hat\sigma^T_{q\bar q\to g+Z^0}}{d\hat t} \simeq
\frac{8 \alpha_s \alpha_{\rm em} c_1^q}{9 \sin^2 2\theta_W} \frac{1}{\hat s} \frac{1}{P_\perp^2} 2(1-\xi) = \sigma_0^q \frac{1}{P_\perp^2} \frac{\alpha_s}{2\pi} C_F 2(1-\xi)
\end{align}
where $\sigma_0^q = \frac{4\pi \alpha_s \alpha_{\rm em} c_1^q}{3 \sin^2 2\theta_W} \frac{1}{\hat s}$ is the leading-order partonic cross section for $q\bar q\to Z^0$ and $C_F = 4/3$. Including the Jacobian factor $1/\xi$ from the phase space transformation $\int dy_{\rm jet} d^2 P_\perp = \int \frac{d\xi}{\xi} d^2P_\perp$, we recover the expressions utilized in our paper. The asymptotic expression for the gluon linear polarization is then only a function of $\xi$ given by
\begin{align}
P_{g, {\rm asymptotic}}^T (\xi) = \frac{2(1-\xi)}{1+(1-\xi)^2}. 
\label{eq:gluon_pol_asy}
\end{align}
We emphasize that the factorized expression serves primarily to elucidate the physical origin of the linear polarization. It reveals that the polarization arises directly from the helicity structure of the collinear splitting functions, establishing it as a fundamental property of the radiation process itself. In a sophisticated phenomenological study, one can always replace the asymptotic expression with the exact one.

\begin{figure}[htb]
\centering
\includegraphics[width=0.5\linewidth]{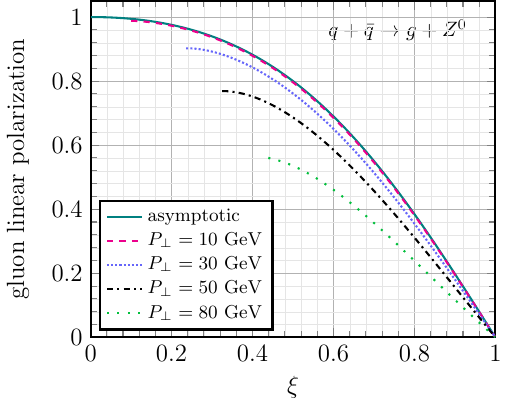}
\caption{Gluon linear polarization of the $q + \bar q \to g +Z^0$ channel as a function of $\xi$ at different $P_\perp$.}
\label{fig:gluon_polarization_pperp}
\end{figure}

In Fig.~\ref{fig:gluon_polarization_pperp}, we compare the linear polarization calculated using the exact result in Eq.~(\ref{eq:gluon_pol}) with the asymptotic expression in Eq.~(\ref{eq:gluon_pol_asy}). At low transverse momentum (e.g., $P_\perp = 10$ GeV), the exact result almost overlaps with the asymptotic curve. As $P_\perp$ increases, deviations arising from finite-$P_\perp$ power corrections (scaling as $P_\perp^2/M_Z^2$) become visible. However, even at $P_\perp \simeq 30$ GeV, the deviation remains small. The asymptotic expression is still a good approximation. Although the modification becomes significant at very large $P_\perp$ (e.g., 80 GeV), the fundamental mechanism generating polarization via parton splitting remains robust, and the magnitude of the gluon linear polarization remains substantial.

\begin{figure}[htb]
\centering
\includegraphics[width=0.5\linewidth]{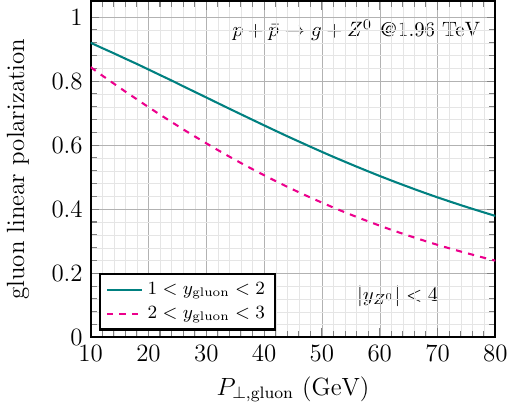}
\caption{Linear polarization of gluons produced in $p\bar p$ collisions at $\sqrt{S}=1.96$ TeV as a function of transverse momentum $P_\perp$.}
\label{fig:gluon_pol_tevatron}
\end{figure}

Furthermore, to validate this in a realistic experimental setup, we computed the gluon linear polarization utilizing the exact expression in Eq.~(\ref{eq:gluon_pol}) for $p\bar p$ collisions at the Tevatron energy ($\sqrt{S}=1.96$ TeV). These results are shown in Fig.~\ref{fig:gluon_pol_tevatron} as a function of $P_\perp$. We required the gluon jet to be in the forward rapidity bins $y_{\rm gluon} \in [1,2]$ and $[2,3]$, integrating the $Z^0$ rapidity over $y_{Z^0} \in [-4,4]$. The numerical results confirm that for the relevant kinematics ($P_\perp \sim 30$ GeV), the asymptotic expression provides a reliable description of the gluon polarization, and the magnitude of the effect is sizeable.

\section{EEC in $e^+e^-$ annihilations}

Energy-energy correlations (EECs) were first proposed for measurement in electron-positron colliders as a clean and important test of QCD \cite{Basham:1978bw, Basham:1978zq}. Over the past several decades, a wealth of experimental data \cite{Schlatter:1981aw, CELLO:1982rca, Fernandez:1984db, TASSO:1987mcs, TOPAZ:1989yod, DELPHI:1990sof, OPAL:1991uui, Electron-PositronAlliance:2025fhk} has been accumulated across a broad range of center-of-mass energies, providing a unique platform for studying the transition between perturbative and non-perturbative QCD dynamics. 

In $e^+e^-$ annihilation, the leading order process $e^+e^- \rightarrow q\bar{q}$ produces two nearly back-to-back jets. Therefore, the near-side EEC mainly probes the energy distribution within the quark-initiated jets, which can be computed from the derivative of the jet function $J_q$, whose evolution is governed by Eq.~(8) in the DGLAP formalism in the main text. In the CCFM formalism, the evolution equation follows one akin to Eq.~(14) by properly changing the parton flavors. For completeness, we lay out the full evolution equation taking into account the off-diagonal terms in the CCFM formalism as follows,
\begin{align}
&
\frac{\partial}{\partial \ln\mu^2} \frac{J_g (\ln \kappa)}{\Delta_{s}(\mu^2)} 
= \frac{\alpha_s}{2\pi} \frac{1}{\Delta_s(\mu^2)} \int_{\Lambda/\mu}^{1-\Lambda/\mu} dy y^2 
\left[ \tilde P_{gg} (y) J_g (\ln\kappa)+ \tilde P_{gq} (y) J_q (\ln\kappa) \right], 
\\
&
\frac{\partial}{\partial \ln\mu^2} \frac{J_q (\ln \kappa)}{\Delta_{s,q}(\mu^2)} 
= \frac{\alpha_s}{2\pi} \frac{1}{\Delta_{s,q}(\mu^2)} \int_{\Lambda/\mu}^{1-\Lambda/\mu} dy y^2 
\left[ \tilde P_{qq} (y) J_q (\ln\kappa) + \tilde P_{qg} (y) J_g (\ln\kappa)\right].
\end{align}
Here, $\Delta_s (\mu^2)$ is the Sudakov form factor for the gluon case which has already been defined in Eq.~(15) in the main text, and $\Delta_{s,q}$ is that for the quark. They read
\begin{align}
&
\Delta_{s} (\mu^2) = \exp\Big\{ - \int_{4\Lambda^2}^{\mu^2} \frac{d\mu'^2}{\mu'^2} \frac{\alpha_s(\mu'^2)}{4\pi} \int_{\Lambda/\mu'}^{1-\Lambda/\mu'} dy \left[\tilde P_{gg} (y) + 2n_f P_{qg} (y) \right] \Big\},
\\
&
\Delta_{s,q} (\mu^2) = \exp\Big\{ - \int_{4\Lambda^2}^{\mu^2} \frac{d\mu'^2}{\mu'^2} \frac{\alpha_s(\mu'^2)}{2\pi} \int_{\Lambda/\mu'}^{1-\Lambda/\mu'} dy \tilde P_{qq} (y) \Big\}.
\end{align}
These unregularized splitting functions are given as
\begin{align}
&
\tilde P_{gg} (y) = 2N_c \left[\frac{1-y}{y} + y(1-y) + \frac{y}{1-y}\right],
\\
&
\tilde P_{qg} (y) = \frac{1}{2} [y^2 + (1-y)^2],
\\
&
\tilde P_{qq} (y) = C_F \frac{1+y^2}{1-y},
\\
&
\tilde P_{gq} (y) = C_F \frac{1+(1-y)^2}{y}.
\end{align}
While solving the evolution equation, the boundary conditions at the reference scale $\mu_0$ (corresponding to the large-angle limit $\theta_0 \sim 0.3$) are specified as $J_q(\ln \kappa = 0) = 1$ and $J_g(\ln \kappa = 0) = 0$, reflecting the fact that the primary partons are quarks and antiquarks.

\begin{figure}[htb]\centering
\includegraphics[width=1\textwidth]{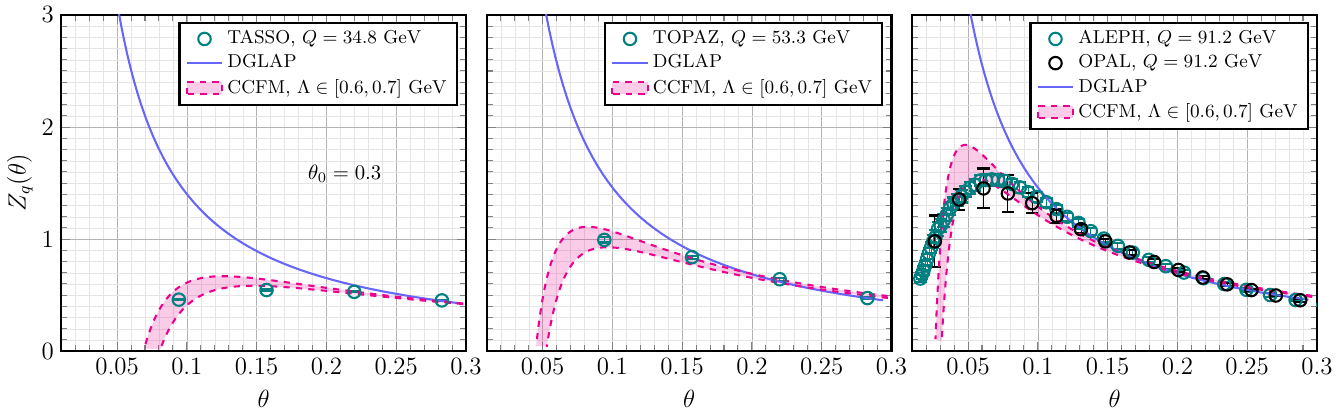}
\caption{Comparison of theoretical calculations for EEC in electron-positron annihilation, calculated within the DGLAP and CCFM formalisms, with experimental data from the TASSO \cite{TASSO:1987mcs}, TOPAZ \cite{TOPAZ:1989yod}, OPAL \cite{OPAL:1991uui} and ALEPH \cite{Electron-PositronAlliance:2025fhk} collaborations. The data points have been rescaled to be comparable with each other at $\theta_0=0.3$. Notice that the ALEPH data are provided by Electron-Positron Alliance \cite{Electron-PositronAlliance:2025fhk} utilizing the Archival ALEPH data.}
\label{fig:eec_in_ee}
\end{figure}

We compare our theoretical calculations, performed within both the DGLAP and CCFM formalisms, with experimental data from TASSO ($Q = 34.8$ GeV) \cite{TASSO:1987mcs}, TOPAZ ($Q = 53.3$ GeV) \cite{TOPAZ:1989yod}, and OPAL and ALEPH (both at $Q = 91.2$ GeV) \cite{OPAL:1991uui, Electron-PositronAlliance:2025fhk}. The comparison is shown in Fig.~\ref{fig:eec_in_ee}. Since different collaborations have adopted different conventions for presenting EEC, we have converted all data to $Z_q(\theta) = d\Sigma/d\theta$ as a function of $\theta$. The data point at $\theta=0$ has been excluded due to the contamination of self-correlations. Furthermore, all data are rescaled to be comparable with each other at $\theta_0 = 0.3$, which allows for a direct comparison of the shape and evolution behavior on different energy scales. 

As shown in Fig.~\ref{fig:eec_in_ee}, a prominent feature of the EEC in $e^+e^-$ collisions is the emergence of a plateau in the small-$\theta$ region. The height of this plateau increases systematically with the collision energy $Q$, and consequently with the initiating quark energy. Physically, this plateau signifies the onset of confinement and the transition into the non-perturbative regime. It has been a subject of intense investigation in the most recent literature, see, e.g., Refs.~\cite{Lee:2024esz, Lee:2025okn, Chang:2025kgq, Herrmann:2025fqy, Kang:2025zto, Guo:2025qnz}, exploring the ``confinement transition'' and the emergence of a scaling behavior \cite{Chang:2025kgq} beyond the perturbative regime. These works highlight that the transition is marked by a clear turning point where the perturbative power-law behavior gives way to a flat plateau. 

The standard DGLAP formalism results in an unphysical power-law divergence at small angles, and fails to describe the observed plateau in the transition region. In contrast, this unphysical increase is effectively tamed in the CCFM formalism by incorporating coherence effects through angular ordering, alongside an infrared cutoff parameter $\Lambda$ that can be fixed by fitting to experimental data.

In conclusion, the plateau behavior is naturally captured within the CCFM framework, leading to a finite and stable distribution as $\theta \to 0$. It thus provides a reliable description of the intermediate transition region and offers a theoretically consistent bridge between the high-energy partonic shower and the low-energy hadronic state, a feature that remains a challenge for traditional collinear factorization.

\section{Fixed order calculation for EEC between $c$ and $\bar c$ in $p\bar p$ collisions}

In this section, we present a fixed-order perturbative calculation of the azimuthal-angle-dependent EEC for $c\bar{c}$ pairs within gluon-initiated jets in $p\bar{p}$ collisions at $\sqrt{S} = 1.96$ TeV, corresponding to Tevatron kinematics. The primary objective is to demonstrate the efficacy of heavy-flavor tagging in enhancing the gluon polarization signal. To this end, we further compare the azimuthal modulation observed in $Z^0$-tagged jet events with that in single inclusive jet production, thereby isolating and quantifying the $\cos 2\phi$ asymmetry induced by linearly polarized gluons.

\begin{figure}[htb]
\centering
\includegraphics[width=0.25\textwidth]{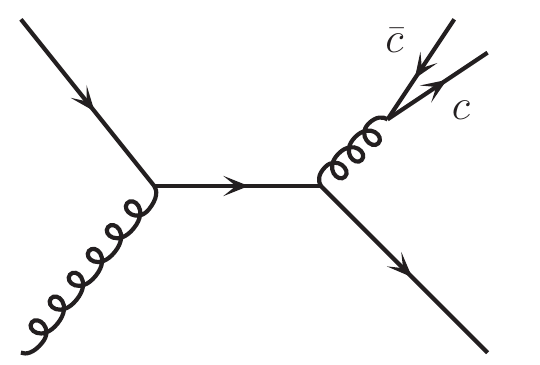}
\qquad
\includegraphics[width=0.25\textwidth]{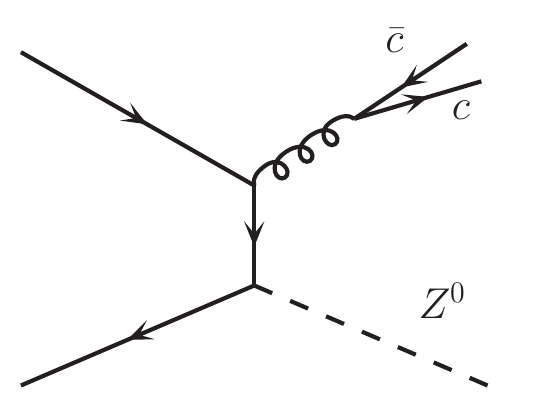}
\caption{Representative leading-order Feynman diagrams for the production of a $c\bar{c}$ pair within a jet in hadronic collisions. The left panel is for the inclusive jet production, while the right panel is for the $Z^0$-tagged jet production.}
\label{fig:ccbar_azimuthal}
\end{figure}

We emphasize that the current resummation formalism is formulated in the massless limit and has not yet been extended to incorporate heavy-quark tagging or mass effects. Consequently, we employ a fixed-order perturbative approach for the analysis of the tagged charm-pair observable. The generalization of the resummation framework to include heavy-quark mass corrections and flavor tagging represents a significant theoretical undertaking left for future work. Given that the $c\bar{c}$ pair is collinear within a single jet, the dominant contribution arises from tree-level $2\to 3$ scattering processes, as depicted in Fig.~\ref{fig:ccbar_azimuthal}. The relevant matrix elements have been derived in Refs.~\cite{Ellis:1985er, Campbell:2000bg} and are implemented in the MCFM program.

\begin{figure}[htb]
\centering
\includegraphics[width=0.5\textwidth]{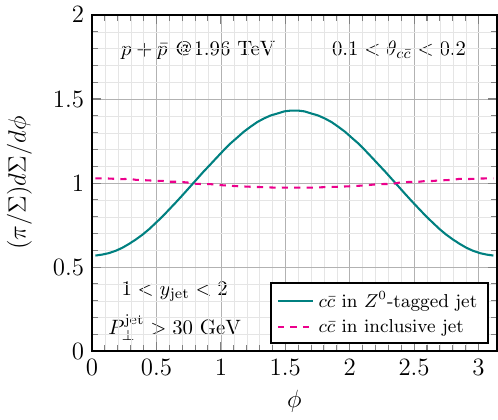}
\caption{Normalized azimuthal angle dependent  heavy flavor tagged  EEC in $p \bar p$ collisions at $\sqrt{S}=1.96$ TeV.}
\label{fig:ccb_numerical}
\end{figure}

We present the numerical results in Fig.~\ref{fig:ccb_numerical}, computed using the Monte-Carlo method developed in Refs.~\cite{Harris:2001sx, Owens:2001rr}. The analysis selects jets containing a $c\bar{c}$ pair in the forward rapidity region (proton-going direction) with jet transverse momentum $P_\perp^{\rm jet} > 30$ GeV. To minimize non-perturbative contamination, the opening angle $\theta_{c\bar{c}}$ between the charm and anti-charm quarks is restricted to the range $[0.1, 0.2]$. The EEC is analyzed as a function of the azimuthal angle $\phi$, defined as the angle between the $c\bar{c}$ plane and the beam-jet plane. For clarity in visualizing the anisotropy, the azimuthal distribution is normalized such that its integral equals $\pi$.  In $Z^0$-tagged jet events, the azimuthal dependence exhibits a pronounced $\cos 2\phi$ modulation with an amplitude reaching approximately 40\%. This is a significant fraction of the theoretical maximum asymmetry of $\sim 60\%$, which corresponds to the $g \to c\bar{c}$ splitting of a maximally linearly polarized gluon.

Conversely, inclusive jet production exhibits negligible azimuthal anisotropy. This marked disparity stems from the distinct helicity structures of the underlying hard processes. Since linearly polarized gluons correspond to an interference between distinct helicity states ($+1$ and $-1$), their generation requires a mechanism capable of flipping helicity. In $Z^0$-associated production, the large mass of the $Z^0$-boson facilitates this helicity flip for the radiated gluon, leading to significant linear polarization and a correspondingly large $\cos 2\phi$ asymmetry in the $c\bar{c}$ pair distribution. In contrast, inclusive jet production is dominated by massless QCD scattering where such helicity flips are forbidden, thereby suppressing the polarization signal.

We conclude that heavy-flavor tagging offers a decisive advantage for gluon polarimetry. By isolating the $g \to c\bar{c}$ or $g\to b\bar b$ splitting channel, we effectively disentangle it from the competing $g \to gg$ contribution. This separation eliminates the partial cancellation that typically dilutes the azimuthal asymmetry in inclusive measurements, thereby yielding a significantly enhanced $\cos 2\phi$ signal. Heavy-flavor tagging thus provides a highly sensitive probe of the underlying gluon polarization and the dynamics of parton splitting. We plan to develop a comprehensive resummation framework for this observable in future work, incorporating the track function formalism to account for flavor tagging effects.

\end{document}